# INTERPLAY BETWEEN THE RAY AND MODE EFFECTS IN ELECTROMAGNETIC BEHAVIOR OF SMALL-SIZE HEMIELLIPTIC DIELECTRIC LENSES

**A.V. Boriskin [1], R. Sauleau [2], A.I. Nosich [1]**

[1] Institute of Radiophysics and Electronics NASU, Acad. Proskury Str. 12, 61085 Kharkiv, Ukraine
[2] Institut d'Electronique et de Télécommunications de Rennes, Université de Rennes 1, 35042 Rennes, France
e-mail a_boriskin@yahoo.com

*Abstract* – Focusing and resonance properties of two-dimensional (2-D) small-size hemielliptic lenses made of different materials are studied numerically in order to estimate the influence of internal reflections on the radiation characteristics of dielectric lens antennas (DLA). Accuracy of in-house made algorithms based on combination of geometrical and physical optics (GO, PO) and FDTD in the analysis of optical and modal effects in the behavior of such lenses is tested by comparison with the exact solution obtained using the Muller boundary integral equations (MBIE). The range of applicability for the approaches is discussed.

## I. INTRODUCTION

Dielectric lenses of various shapes, sizes, and made of various materials have become a constitutive part of many novel systems operating at millimeter (mm), sub-mm, and THz ranges including communication systems, radars, imaging and spectroscopy systems, etc. [1-2]. They are commonly used to improve radiation characteristics of the primary feeds (i.e. metallic waveguides, horns or printed antennas) or, reciprocally, to enhance sensitivity of detectors placed in the foci of the lenses. Direct integration of the feeds with dielectric lenses of specific shapes has enabled one to design compact antennas with desired radiation patterns. In contrast to the use of lenses in optics where lenses are put in the far zone of the light emitters, integration technology provides a number of additional benefits that make integrated DLAs very attractive solution for a wide range of applications, namely size, weight, and cost reduction, improved matching, possibility of mass production, etc.

As it has been demonstrated recently, dielectric lenses as small as several wavelengths in free space can already provide a significant improvement of the radiation characteristics of printed antennas [2]. Later, DLAs with reduced-size shaped lenses, whose profiles have been optimized to provide desired radiation patterns, have been introduced [3]. Although a reasonable agreement between theoretical estimations and experimental data was obtained, these studies highlighted the deficiency of the available simulation tools when applied to the analysis and optimization of such DLAs. A solver for a CAD tool, suitable for efficient and reliable design and optimization of DLAs, must be fast, have reasonable requirements to the computational resources, and guarantee a controlled accuracy for any set of antenna parameters. For finite-size arbitrary-shaped lenses this is a challenging problem because each dielectric lens, in fact, is an open dielectric resonator of specific shape. Thus its electromagnetic behavior is determined by the interplay of two major mechanisms, namely, geometrical-optics focusing and wavelength-scale internal resonances. The latter can be excited if the incident field frequency hits the real part of the complex-valued frequency of a sufficiently high-Q natural mode of the resonator. Out of the high-Q resonances, geometrical-optics focusing is dominant. Therefore, low-index lenses ($\varepsilon \leq 4$) typically used in the integrated DLAs operating at the mm and sub-mm wavelength ranges, generally demonstrate the priority of the focusing mechanism over the resonance one. However, the latter can become dominant for small-size lenses made of high-index material ($\varepsilon \geq 10$) such as silicon [4]. This co-existence and overlap of the optical and modal features in the behaviors of small-size dielectric lenses makes accurate description of their electromagnetic properties a challenging task. As it was recently demonstrated, the high-frequency approximations, such as GO and PO, are not applicable for the description of the focusing properties of such lenses [5] even though these methods are often successfully used for the analysis of DLAs in the emitting mode [3, 6-8]. This is because they fail to account for the curvature of the lenses and hence accurately reproduce the multiple internal reflections even for the off-resonance excitation. Accurate description of the resonances in such lenses made of dense materials is beyond accessibility of both GO and PO; furthermore, it can become troublesome even for full-wave approaches, such as FDTD [9].

In the paper, we summarize the results of the studies aimed at (i) precise description of electromagnetic properties of reduced-size dielectric lenses and (ii) assessment of the accuracy of numerical algorithms based on GO-PO and FDTD approaches in the analysis of small-size lenses. We consider the problem in two-dimensional (2D) formulation and obtain the trusted reference solution by using a highly efficient numerical algorithm based on the combination of the MBIE, analytical regularization, and trigonometric Galerkin discretization scheme [10]. This approach guarantees full account for the finite size of the lens as well as its curvature. The parameters of the lenses considered correspond to off-resonance and resonance excitations. The influence of the internal resonances on the focusing and collimating properties of the lenses is discussed.

## II. PROBLEM FORMULATION AND OUTLINE OF SOLUTION

The lens is considered as a homogeneous dielectric cylinder whose profile is described by a smooth analytical curve, which consists of a a hemiellipse and a hemi-super-ellipse (rectangle with rounded corners) joint at the points $(0, \pm a)$, where $a$ is the minor semi-axis of the ellipse. Eccentricity of the hemielliptic part is chosen in accordance to the GO focusing rule, i.e. $e = \varepsilon^{-1/2}$, and the lens extension size equals the ellipse focal distance, $al = f$. For numerical simulations we consider lenses made of rexolite ($\varepsilon = 2.53$), quartz ($\varepsilon = 3.8$), and silicon ($\varepsilon = 11.7$), and vary the bottom size $2a$ between 3 and 4 wavelengths in free space, $\lambda_0$. In figures, all lens dimensions are normalized by the free space wavenumber, $k = 2\pi/\lambda_0$.

The algorithms used for the analysis are based on three different approaches, namely a combination of GO-PO, FDTD, and MBIE. The latter is used as a reference solution thanks to its controllable accuracy for any set of lens parameters [6, 9, 10]. The algorithms under test were developed in the following manner. The GO-PO algorithm implements a two-step procedure: (i) GO ray-tracing is used to compute the equivalent electromagnetic currents flowing on the dielectric boundary of the lens and (ii) the near fields are derived from the so-called PO formulation based on the Huygens-Kirchhoff principle for the 2-D space [3]. The upside of this approach is simplicity and low time and memory consumption, the downside is neglection of the lens profile curvature and thus uncontrolled error. In-house FDTD-2D solver was developed in line with the standard method with Cartesian grid. FDTD-based CAD-tools are known to be very flexible and user-friendly. This often leaves in the shadow the bottleneck of this approach, i.e. the failure to characterize high-Q resonances [11]. Unlike approximate methods, MBIE-Galerkin algorithm is equally accurate and economic off resonances and in their vicinities.

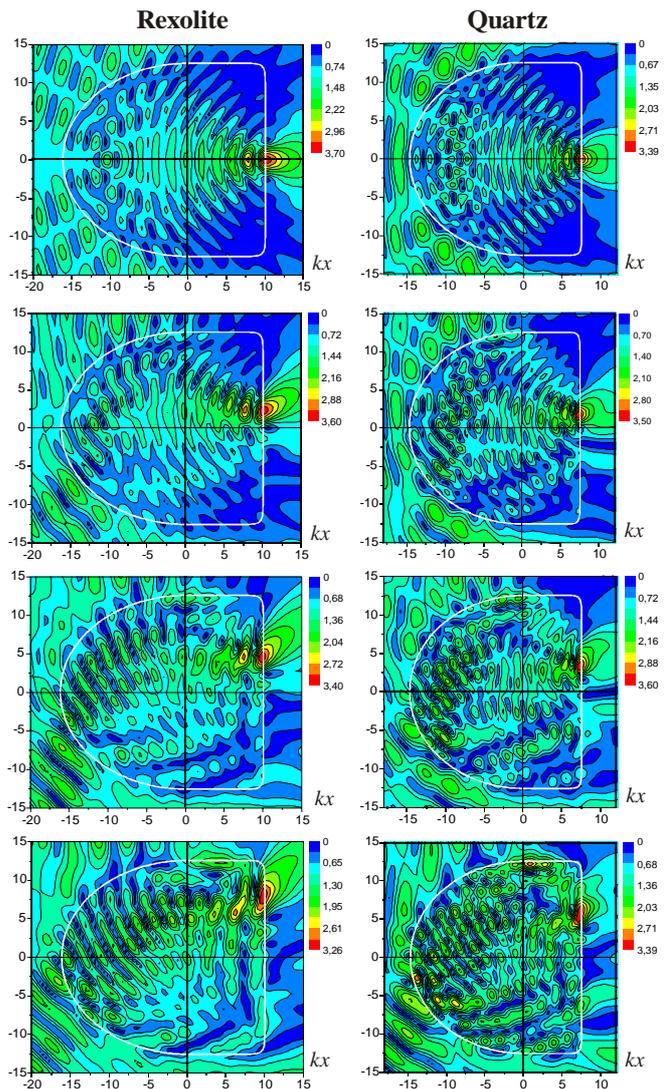

Fig. 1. Near-field maps for the hemielliptic lenses ($a = 2\lambda_0$) illuminated by the unit-amplitude plane $E$-wave under 0°, 10°, 20°, and 30° angles of incidence.

## III. RESULTS AND DISCUSSION

The exact near-field maps computed by MBIEs for the rexolite and quartz lenses illuminated by the unit-amplitude *E*-polarized plane waves (Fig. 1) clearly demonstrate the important role of internal reflections in the patterns formation. Here, in contrast to the GO prediction of a single focal point located in the ellipse rear focus, e.i. right at the bottom, standing wave patterns with several spots of almost equal field amplitudes are observed. For the oblique incidence ($\gamma \neq 0°$) the spots migrate along the flat bottom of the lens and vary both in size and shape. In order to visualize the field intensity distribution along the lens bottom and to test the applicability of high-frequency approaches for analyses of the focusing properties of the small-size lenses the corresponding curves have been computed by MBIE and GO-PO algorithms (Fig. 2). Although the lenses of the selected size, shape, and material do not support high-Q resonances, the GO-PO approach is not capable for accurate description of the near-fields in the focal area, i.e. it enables one to predict only the location of the focal spots but fails to characterize the size and the level of side-spots. Account of multiple reflections slightly improves the accuracy but still does not compensate the inaccuracies entering the solution due to neglection of the lens curvature when computing the ray trajectories within the lens [5].

Although DLAs are often considered as quasi-optical devices, which is intuitively dictated by the optical associations, such antennas, especially those having reduced-size lenses made of dense material like silicon, superpose both ray-type and mode-type behaviors. To demonstrate this and to test accuracy of the in-house FDTD-2D algorithm, we plotted the main-beam directivity of a 2D hemielliptic silicon lens excited by electrical and magnetic currents vs. the lens extension parameter (Fig. 3). In such a way, we find the so-called half-bowtie resonances (HBT) [12] seen as deep periodic drops in forward directivity. Comparison of the curves obtained by two methods shows that the FDTD algorithm displays a regular shift of the whole curve which become evident near the resonance frequencies. This shift is explained in part by the staircasting error in the discretization of the curved contour and by the reflections from the borders of the computational window equipped with a perfectly matching layer (PML). Although the PML parameters were chosen to guarantee the -50 dB normal back reflection, for other angles of incidence this value can be much higher and corners of the computational window usually "shine" quite significantly. To visualize the HBT resonance and to describe its influence on the collimating properties of the 2D lens, a near-field map (Fig. 4) and normalized radiation patterns (Fig. 5) have been computed for the lens whose parameters corresponded to the resonance indicated by a triangle in Fig. 3. For comparison, radiation patterns are plotted for the lenses with two different extensions, namely the one cut through the rear focus (indicated by the vertical dashed line in Fig. 3) and the resonance one ($l_1 = 0.36$). As to the accuracy of the FDTD-2D solver for lenses of given size, shape, and material, it is found to be reasonable for both near- and far-field characteristics [10].

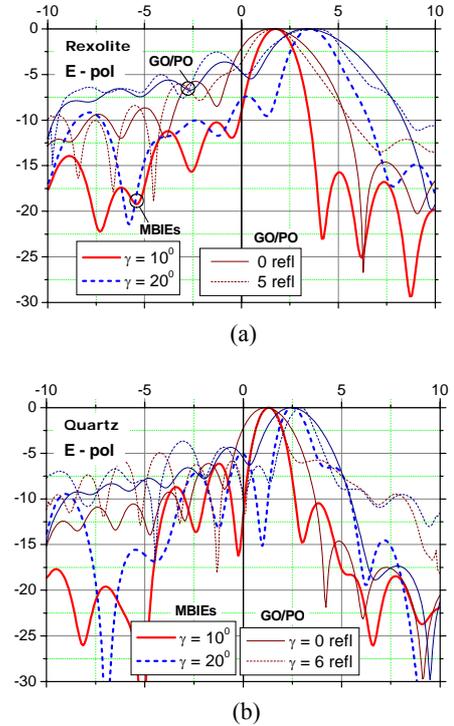

Fig. 2. Normalized field intensity inside the lenses in the plane parallel to the lens bottom. The lenses are illuminated by *E*-plane wave in a grazing mode. The GO/PO curves are obtained with and w/o accounting for internal reflections.

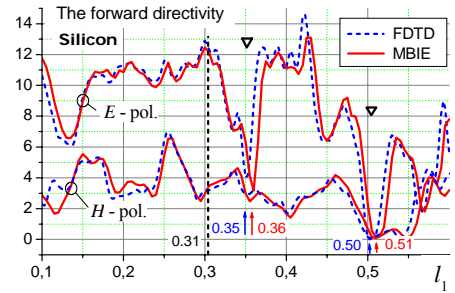

Fig. 3. The forward directivity of electrical and magnetic line currents exciting a hemielliptic silicon lens ($ka = 9.42$) vs. the lens extension parameter. The vertical dashed line indicates the location of the rear focus of the ellipse.

## IV. CONCLUSION

The ray-like and modal properties of small-size hemielliptic lenses, typically used as building blocks of integrated DLAs operating at the mm and sub-mm wavelength ranges, have been described accurately by the MBIE. Moreover, comparison of the exact MBIE solution with those obtained by the in-house algorithms based on GO-PO and FDTD-2D has enabled us to estimate the accuracy of these approaches as to the analysis of focusing and collimating abilities of such lenses.

### ACKNOWLEDGEMENT

This work was supported in part by the joint projects between the Institute of Radiophysics and Electronics of the National Academy of Sciences of Ukraine, on the one side, and the Centre National de la Recherche Scientifique, Ministère de l'Education Nationale, de l'Enseignement Supérieur et de la Recherche, and Ministère des Affaires Etrangères et Européennes, France, on the other side. The first author was also supported by the Foundation Michel Métivier, by the Brittany Region via the CREATE/CONFOCAL project, and by the NATO Reintegration Grant NUKR.RIG.983313.

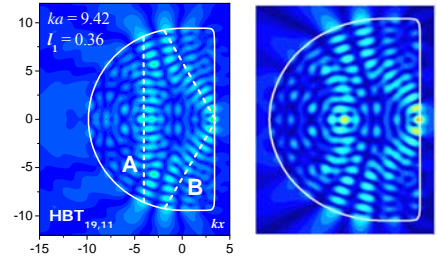

Fig. 4. Normalized near-field intensity maps of extended hemielliptic silicon lenses ($ka = 9.42$) excited by electric line currents. The lens extension value corresponds to the resonance indicated by the triangle in Fig. 3. Solid white line is for the lens cross-section; dashed line highlights the characteristic triangular shape of the HBT resonance.

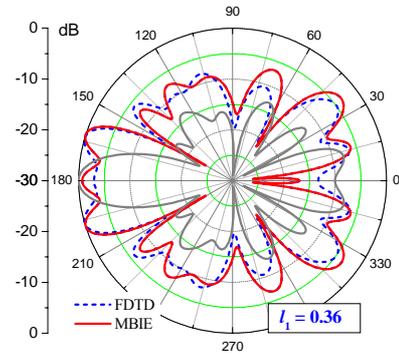

Fig. 5. The normalized radiation patterns of the electric line current illuminating silicon hemielliptic lenses with resonance extension marked by the triangle in Fig. 3. For comparison, grey line is for the radiation pattern of the line source illuminating the lens cut through the rear GO focus ($l_1 = 0.31$) computed by MBIEs.

### REFERENCES

[1.] L.C. Godara, *Handbook of Antennas in Wireless Communications*, Ch. 15 by C. Fernandes, CRC Press, 2002.
[2.] G. Godi, R. Sauleau, and D. Thouroude, "Performance of reduced size substrate lens antennas for millimetre-wave communications," *IEEE Trans. Antennas Propag.*, vol. 53, no. 4, pp. 1278-1286, 2005.
[3.] B. Barès, R. Sauleau, L. Le Coq, and K. Mahdjoubi, "A new accurate design method for millimeter-wave homogeneous dielectric substrate lens antennas of arbitrary shape," *IEEE Trans. Antennas Propag.*, vol. 53, no. 3, pp. 1069-1082, 2005.
[4.] A.V. Boriskin, S.V. Boriskina, A.I. Nosich, T.M. Benson, P. Sewell, and A. Altintas, "Lens or resonator? – electromagnetic behavior of an extended hemielliptical lens for a sub-mm wave receiver," *Microw. Opt. Tech. Lett.*, vol. 43, no. 6, pp. 515-158, 2004.
[5.] A.V. Boriskin, G. Godi, R. Sauleau, and A.I. Nosich, "Small hemielliptic dielectric lens antenna analysis in 2-D: boundary integral equations vs. geometrical and physical optics," *IEEE Trans. Antennas Propag.*, vol. 56, no. 2, pp. 485-492, 2008.
[6.] X. Wu, G. V. Eleftheriades, and T.E. van Deventer-Perkins, "Design and characterization of single- and multiple-beam mm-wave circularly polarized substrate lens antennas for wireless communications," *IEEE Trans. Microw. Theory Tech.*, vol. 49, no. 3, pp. 431-441, 2001.
[7.] D. Pasqualini and S. Maci, "High-frequency analysis of integrated dielectric lens antennas," *IEEE Trans. Antennas Propag.*, vol. 52, no. 3, pp. 840-847, 2004.
[8.] R. Sauleau and B. Barès, "A complete procedure for the design and optimization of arbitrarily shaped integrated lens antennas," *IEEE Trans. Antennas Propag.*, vol. 54, no. 4, pp. 1122-1133, 2006.
[9.] A.V. Boriskin, A. Rolland, R. Sauleau, and A.I. Nosich, "Assessment of FDTD accuracy in the compact hemielliptic dielectric lens antenna analysis," *IEEE Trans. Antennas Propag.*, vol. 56, no. 3, pp. 758-764, 2008.
[10.] S.V. Boriskina, T.M. Benson, P. Sewell, and A.I. Nosich, "Accurate simulation of 2D optical microcavities with uniquely solvable boundary integral equations and trigonometric-Galerkin discretization," *J. Opt. Soc. Am. A*, vol. 21, no. 3, pp. 393-402, 2004.
[11.] A.V. Boriskin, A. Rolland, R. Sauleau, and A.I. Nosich, "Test of the FDTD accuracy in the analysis of the scattering resonances associated with high-Q whispering-gallery modes of a circular cylinder," *J. Opt. Soc. Am. A*, vol. 25, no. 5, pp. 1169-1173, 2008.
[12.] J. Wiersig, "Formation of long-lived, scarlike modes near avoided resonance crossings in optical microcavities," *Phys. Rev. Lett.*, vol. 97, 253901, 2006.